\documentclass[a4paper,12pt]{article}
\usepackage{epsf,amsmath}
\usepackage{graphicx}
\usepackage{amssymb}
\usepackage{array,tabularx}
\usepackage{indentfirst}
\usepackage[utf8]{inputenc}
\usepackage{cite}
\begin{document}

\title{Dynamical stability of extended teleparallel gravity.}

\author{Petr V. Tretyakov}
\date{}
\maketitle
{\it \begin{center}Bogoliubov Laboratory of Theoretical Physics, Joint Institute for Nuclear Research\end{center}
\begin{center}Joliot-Curie 6, 141980 Dubna, Moscow region, Russia\end{center}
\begin{center}tpv@theor.jinr.ru\end{center}}

\begin{abstract}
We discuss modified teleparallel gravity with function $f(T,T_G)$ in the action, where function depend on two arguments: torsion scalar $T$ and analogue of Gauss-Bonnet invariant $T_G$. In contradistinction to usual teleparallel gravity $f(T)$, this theory contains higher derivative terms, which may produce different instabilities. We discuss Minkowski stability problem in such kind of theories and explicitly demonstrate that for stability must be $f_T(0,0)<0$, $f_{T_GT_G}>0$. We apply these restrictions for the few types of functions discussed by the early authors.

\vspace{0.5cm}
PACS: 04.05.Kd

\vspace{0.5cm}
Keywords: Modified gravity; Gauss-Bonnet; $f(T)$ gravity; dynamical stability.

\end{abstract}

\section{Introduction}

Teleparallel gravity \cite{AP,Maluf} is one of the possible modifications of gravity, which give us possibility to solve a wide class of cosmological problems \cite{BMNO,BHO,OS,BOS,BNO,MM,SST,SSU,CCLS}. Orthonormal tetrad components $e_A(x^{\mu})$, or so-called vierbein, are used for formulation of teleparallel gravity. The relation to the metric takes the next form $g_{\mu\nu}=\eta_{AB}e^A_{\mu}e^B_{\nu}$, where Latin indices runs over $0,1,2,3$ for the tangent space at each point $x^{\mu}$ of the manifold, Greek one are coordinate indices on the manifold, which also run over $0,1,2,3$ and $e^A_{\mu}$ forms the tangent vector of the manifold. The torsion $T^{\lambda}_{\,\,\,\,\mu\nu}$ and contorsion $K^{\mu\nu}_{\,\,\,\,\,\,\,\,\lambda}$ tensors are defined as
\begin{equation}
T^{\lambda}_{\,\,\,\,\mu\nu}\equiv e^{\lambda}_A(\partial_{\mu}e^A_{\nu}-\partial_{\nu}e^A_{\mu}),
\label{0.1}
\end{equation}
\begin{equation}
K^{\mu\nu}_{\,\,\,\,\,\,\,\,\lambda}\equiv -\frac{1}{2}\left ( T^{\mu\nu}_{\,\,\,\,\,\,\,\,\lambda} -T^{\nu\mu}_{\,\,\,\,\,\,\,\,\lambda} -T_{\lambda}^{\,\,\,\,\mu\nu} \right ),
\label{0.2}
\end{equation}
and the torsion scalar $T$ is
\begin{equation}
T\equiv \frac{1}{4}T^{\lambda\mu\nu}T_{\lambda\mu\nu}+\frac{1}{2}T^{\lambda\mu\nu}T_{\nu\mu\lambda}- T_{\lambda\mu}^{\,\,\,\,\,\,\,\,\lambda}T^{\nu\mu}_{\,\,\,\,\,\,\,\,\nu}.
\label{0.3}
\end{equation}
Now generalized teleparallel gravity may be written as \cite{FF,BF}
\begin{equation}
S=\frac{1}{2\kappa^2}\int d^4x e f(T),
\label{0.4}
\end{equation}
where $e=det(e^A_{\mu})=\sqrt{-g}$ and $\kappa^2$ is the gravitational constant.

In this paper we discuss extended version of teleparallel gravity with the action defined by
\begin{equation}
S=\frac{1}{2\kappa^2}\int d^4x e f(T,T_{G}),
\label{1.1}
\end{equation}
where $T$ is torsion scalar (\ref{0.3}) and $T_G$ is the teleparallel equivalent of the
Gauss-Bonnet combination $G=R^2+4R^{\mu\nu}R_{\mu\nu}+R^{\mu\nu\rho\lambda}R_{\mu\nu\rho\lambda}$ in such kind formulation of gravity:
\begin{equation}
\begin{array}{r}
T_G=\left ( K^{\kappa}_{\,\,\,\,\phi\pi} K^{\phi\lambda}_{\,\,\,\,\,\,\,\,\rho} K^{\mu}_{\,\,\,\,\xi\sigma} K^{\xi\nu}_{\,\,\,\,\,\,\,\,\tau} - 2K^{\kappa\lambda}_{\,\,\,\,\,\,\,\,\pi}K^{\mu}_{\,\,\,\,\phi\rho}K^{\phi}_{\,\,\,\,\chi\sigma}K^{\chi\nu}_{\,\,\,\,\,\,\,\,\tau}\right. \\
\\ \left. +2K^{\kappa\lambda}_{\,\,\,\,\,\,\,\,\pi}K^{\mu}_{\,\,\,\,\phi\rho}K^{\phi\nu}_{\,\,\,\,\,\,\,\,\xi}K^{\xi}_{\,\,\,\,\sigma\tau} +2K^{\kappa\lambda}_{\,\,\,\,\,\,\,\,\pi}K^{\mu}_{\,\,\,\,\phi\rho}K^{\phi\nu}_{\,\,\,\,\,\,\,\,\sigma,\tau} \right )\delta^{\pi\rho\sigma\tau}_{\kappa\lambda\mu\nu}.
\end{array}
\label{1.1.1}
\end{equation}
Scalar combination $T_G$ in arbitrary dimensions (\ref{1.1.1}) and action (\ref{1.1}) was firstly described in \cite{KS1}. Cosmology in such kind of theories was studied in the few later works \cite{CJMM,KS2,AJ1,AJ2,JRC}. Possible further generalizations of this theory by analogy with Lovelock gravity was proposed in \cite{GV}. Without dipping into details we can take equations of motion for FLRW-metric $ds^2=-dt^2+a^2(t)dx_idx^i$ in the form \cite{KLS}
\begin{equation}
f-12H^2f_T-T_Gf_{T_G}+24H^3\dot f_{T_G}=2\kappa^2\rho_m,
\label{1.2}
\end{equation}
\begin{equation}
f -4(3H^2+\dot H)f_T -4H\dot f_T -T_Gf_{T_G} +\frac{2}{3H}T_G\dot f_{T_G} +8H^2\ddot f_{T_G} = -2\kappa^2p_m,
\label{1.3}
\end{equation}
where $f_{T}$, $f_{T_G}$ denotes partial derivatives of function $f$ with respect to arguments and time derivatives must reads like $\dot f_{T_G}=f_{TT_G}\dot T + f_{T_GT_G}\dot T_G$ and so on. Finally expressions for $T$ and $T_G$ reads for FLRW ansatz as
\begin{equation}
T=6H^2,
\label{1.4}
\end{equation}
\begin{equation}
T_G=24H^2(\dot H+H^2).
\label{1.5}
\end{equation}
The most interesting thing for our study is as follows. In the standard $f(T)$ gravity equations of motion do not contain higher derivatives terms, whereas in the theory (\ref{1.1}) such terms appears (see $\dot f_{T_G}$ combination in Friedman equation (\ref{1.2})). It is well known that a number of different instabilities can arise from this fact. But from another hand this instabilities may be used for finding some fundamental restrictions on the theory. For instance there are restrictions $f_R>0$, $f_{RR}>0$ in $f(R)$ gravity which arise from higher derivatives instabilities. One of such fundamental restrictions may be stability of Minkowski space, which we will discuss in the next section.

\section{Stability conditions}

Let us try to study Minkowski solution stability in the theory (\ref{1.1}). First of all, note that studying stability of Minkowski space (by dynamical system approach) usually very hard task, because corresponding point is degenerated. Nevertheless we may use the next trick, main idea of which was proposed in \cite{T}. We introduce non zero cosmological constant $\Lambda$ in equations and find eigenvalues for the point which appear due to this term. After this we can put $\Lambda\rightarrow +0$ and find asymptotic of eigenvalues. From another hand we know that Minkowski space is stable around us, and this fact allows us restrict parameters of theory: parameters leading to unstable Minkowski space must be excluded from further investigations.

First of all, we need to represent equation (for our case Friedman equation (\ref{1.2}) will enough) in the form of dynamical system

\begin{equation}
\left \{
\begin{array}{l}
\dot H=D,\\
\\ \dot D =F(H,D),
\end{array}
\label{2.1}
\right.
\end{equation}
with
\begin{equation}
\begin{array}{l}
F= \left [\Lambda -f +12H^2f_T + 24H^2(H^2+D)f_{T_G} \right ]\frac{1}{24^2H^5f_{T_GT_G}} \\
\\- \left [ Df_{TT_G} +4D(H^2+D)f_{T_GT_G} \right ]   \frac{1}{2Hf_{T_GT_G}} - 2HD,
\end{array}
\label{2.2}
\end{equation}
where we take into account (\ref{1.4})-(\ref{1.5}). Equation for eigenvalues $\mu$ at some stationary point takes the form
\begin{equation}
\left |
\begin{array}{l}
-\mu\,\,\,\,\,\,\,\,\,\,\,\,\,\,\,\,\,\,\,\,\,\,1\\
\\ (F_H)_0\,\,\,\,\,\,(F_D)_0-\mu
\end{array}
\label{2.3}
\right |=0,
\end{equation}
where $(F_H)_0$, $(F_D)_0$ are values of partial derivatives at the studying point, and its solution
\begin{equation}
\mu_{1,2}=\frac{1}{2}\left[ (F_D)_0 \pm\sqrt{(F_D)_0^2+4(F_H)_0} \right],
\label{2.4}
\end{equation}
tell us that stability conditions of the studying point takes the form\footnote{Remind that $Re(\mu_{1,2})<0$ needs for stability.}
\begin{equation}
(F_D)_0<0,
\label{2.5}
\end{equation}
\begin{equation}
(F_H)_0<0.
\label{2.6}
\end{equation}
After some calculations we can find the next expressions for these values\footnote{Here we took into account that $D_0\equiv0$ for any stationary point.}
\begin{equation}
(F_D)_0=-3H_0,
\label{2.7}
\end{equation}
\begin{equation}
(F_H)_0=\frac{1}{2H_0^4f_{T_GT_G}}\left( f_T + 12H_0^2f_{TT} + 5H_0^4\cdot 24f_{TT_G} +8H_0^6\cdot 24f_{T_GT_G} \right),
\label{2.8}
\end{equation}
where all partial derivatives values must be taken at the studying stationary point $(H_0,0)$. We can see that first stability condition (\ref{2.5}) are always satisfied in expanding universe ($H_0>0$). Second condition (\ref{2.6}) has more complicated structure, but it is easy to see that stability of studying point impossible if all derivatives $f_T$, $f_{TT}$, $f_{TT_G}$, $f_{T_GT_G}$ have similar sign at this point.

Note that up to this point all our reasoning was true for any stationary point of the system (\ref{2.1}), which is actually de Sitter point. Now let us turn to the equation for stationary point and focus our attention on the Minkowski point (which is partial case of de Sitter point). Equation for stationary point takes the form
\begin{equation}
24H_0^4f_{T_G}+12H_0^2f_T+(\Lambda-f)=0,
\label{2.9}
\end{equation}
where all values of functions must be taken at the studying point. First of all note, that we must exclude $f$ from this equation, i.e. put $f(0,0)=0$, otherwise there is no true limit for $H_0$ and even true Minkowski solution will be absent. Equation (\ref{2.9}) has two solutions:
\begin{equation}
H_0^2=\frac{1}{48f_{T_G}}\left [ -12f_T + \sqrt{12^2f_T^2-4\cdot 24f_{T_G}\Lambda} \right], \,\,\, \mathrm{for} \,\,\, f_T>0,
\label{2.10}
\end{equation}
and
\begin{equation}
H_0^2=\frac{1}{48f_{T_G}}\left [ -12f_T - \sqrt{12^2f_T^2-4\cdot 24f_{T_G}\Lambda} \right], \,\,\, \mathrm{for} \,\,\, f_T<0,
\label{2.11}
\end{equation}
because we must study de Sitter point $H_0$ which transform (or tend) to Minkowski point ($H_0\rightarrow +0$) when parameter $\Lambda\rightarrow +0$. Moreover, we can see that there is the only physical solution (\ref{2.11}): for (\ref{2.10}) we have $H_0^2\rightarrow-0$ when $\Lambda\rightarrow +0$ for any sign of $f_{T_G}$, whereas for (\ref{2.11}) contrarily, we have $H_0^2\rightarrow+0$ when $\Lambda\rightarrow +0$ for any sign of $f_{T_G}$. Practically it means that for any theory (\ref{1.1}) must satisfy condition
\begin{equation}
f_T(0,0)<0,
\label{2.12}
\end{equation}
otherwise true vacuum (Minkowski) solution just absent. Now very easy to calculate asymptotic for $H_0$:
\begin{equation}
\lim_{\Lambda\rightarrow +0}H_0^2= \frac{-\Lambda}{12f_T(0,0)}.
\label{2.13}
\end{equation}
Backing to second stability condition (\ref{2.6}) it is easy to find that for functions with $|f_{TT}(0,0)|<\infty$, $|f_{TT_G}(0,0)|<\infty$, $|f_{T_GT_G}(0,0)|<\infty$ must satisfy also
\begin{equation}
f_{T_GT_G}(0,0)>0.
\label{2.14}
\end{equation}

In the next section we discuss some concrete examples of functions, which were proposed by other authors.

\section{Concrete examples of functions}
\subsection{}\label{3a}

Let us start from the simplest function, which were discussed in \cite{KLS}
\begin{equation}
f(T,T_G)=-T+\beta_1\sqrt{T^2+\beta_2T_G}.
\label{3.1}
\end{equation}
First of all note, that first and second derivatives are infinite at $(0,0)$ point, so we cannot use (\ref{2.14}), but need to use a more general one (\ref{2.6}). Note also that near interesting point we have $T\sim 6H_0^2$ and $T_G\sim 24H_0^4$.  Now let us calculate all derivatives from expression (\ref{2.8})
\begin{equation}
(f_{T})_0=-1+\frac{6\beta_1}{\sqrt{36+24\beta_2}},
\label{3.1.1}
\end{equation}
\begin{equation}
(f_{TT})_0=\frac{1}{H_0^2}\frac{\beta_1}{\sqrt{36+24\beta_2}}-\frac{1}{H_0^2}\frac{36\beta_1}{\sqrt{36+24\beta_2}^3},
\label{3.1.2}
\end{equation}
\begin{equation}
(f_{TT_G})_0=\frac{-1}{H_0^4}\frac{3\beta_1\beta_2}{\sqrt{36+24\beta_2}^3},
\label{3.1.3}
\end{equation}
\begin{equation}
(f_{T_GT_G})_0=\frac{-1}{H_0^6}\frac{\beta_1\beta_2^2}{4\sqrt{36+24\beta_2}^3}.
\label{3.2}
\end{equation}
Since we need only signature of expression (\ref{2.8}), we can write
\begin{equation}
(F_H)_0=A\beta_1 \left( 1 + \frac{2\beta_1(\beta_2-3)}{\sqrt{36+24\beta_2}} \right ),
\label{3.3}
\end{equation}
where $A$ some positive constant. Studying of expression (\ref{3.3}) give us the next stability conditions. Minkowski solution stable if
\begin{equation}
-\frac{3}{2}<\beta_2\leqslant3,\,\,\,\beta_1<0,
\label{3.4}
\end{equation}
or
\begin{equation}
-\frac{3}{2}<\beta_2\leqslant3,\,\,\,\beta_1>\frac{\sqrt{6\beta_2+9}}{(3-\beta_2)},
\label{3.4.1}
\end{equation}
or
\begin{equation}
\beta_2>3,\,\,\,-\frac{\sqrt{6\beta_2+9}}{(\beta_2-3)}<\beta_1<0.
\label{3.5}
\end{equation}

\subsection{}

Also there is some modification of function (\ref{3.1})
\begin{equation}
f(T,T_G)=-T+f_1(T^2+\beta_2T_G),
\label{3.6}
\end{equation}
which was discussed in \cite{KS2}. First stability condition (\ref{2.12}) gives us
\begin{equation}
f_T=-1+2f_1'T,
\label{3.7}
\end{equation}
where $'$ denote derivative with respect to full argument of function $f_1$, i.e. $' \equiv\frac{\partial}{\partial x}$, where $x\equiv T^2+\beta_2T_G$. We can see that condition $f_T(0,0)<0$ always satisfied for non-infinite values of $f_1'(0)$. Second condition (\ref{2.14}) gives us
\begin{equation}
f_{T_GT_G}=\beta_2^2 f_1''>0,
\label{3.8}
\end{equation}
so we can see that for Minkowski stability must satisfy $f_1''(0)>0$.  Note that this result will true only for functions with finite values of derivatives at zero point, for functions with infinite values we need to use more complicated procedure described in the previous section \ref{3a}.

\subsection{}

There is another modification of function (\ref{3.1})
\begin{equation}
f(T,T_G)=-T+\beta_1\sqrt{T^2+\beta_2T_G}+\alpha_1T^2+\alpha_2T\sqrt{|T_G|},
\label{3.9}
\end{equation}
which also was discussed in \cite{KS2}. It is clear that first stability condition (\ref{2.12}) give us totally identical result from section \ref{3a}, because all additional terms vanish at the $(0,0)$ point. Moreover second stability condition (\ref{2.14}) give us totally identical result as well, because all additional terms tends to infinity more slowly near $(0,0)$ point. So Minkowski stability conditions for function (\ref{3.9}) absolutely similar as for function (\ref{3.1}).

\subsection{}

Now let us discuss more complicate function, which was obtained firstly in \cite{CJMM} and studied in \cite{AJ1}

\begin{equation}
f(T,T_G)=-T+\sqrt{T}F\left (\frac{T_G}{\sqrt{T}} \right )+\frac{6^{1-s/2}n^2m_p^{4-s}}{s-1}T^{\frac{s}{2}},
\label{3.10}
\end{equation}
where $s$ and $n$ are dimensionless parameters and imply $s\leqslant 2$, $s\neq 0$. Note that near $(0,0)$ point $T\sim 6H^2$, $T_G\sim 24H^4$ and $H\sim 0$, so we must put in our stability conditions $F(0)$. First derivative with respect to $T$ takes the form
\begin{equation}
f_T(0,0)=-1+\frac{F(0)}{2\sqrt{T}}-\frac{T_GF'(0)}{2T}+\frac{s}{2}\frac{6^{1-s/2}n^2m_p^{4-s}}{s-1}T^{\frac{s}{2}-1},
\label{3.11}
\end{equation}
where $'$ denotes derivative with respect to $y\equiv \frac{T_G}{\sqrt{T}}$. Let us discuss functions $F$ with finite values $F(0)$ and $F'(0)$ only. In this case first and third terms in (\ref{3.11}) may be neglected. So first stability condition takes the form
\begin{equation}
\begin{array}{l}
1<s\leqslant 2, \,\,\,\,F(0)<0;\\
\\ 0<s<1, \,\,\,\,{\rm stable\,\, for\,\, any\,\,other\,\, parameters}\\
\\ s<0, \,\,\,\,{\rm unstable\,\, for\,\, any\,\,other\,\, parameters}
\end{array}
\label{3.12}
\end{equation}
Second stability conditions reads
\begin{equation}
f_{T_GT_G}(0,0)=\frac{1}{\sqrt{T}}F''(0)>0,
\label{3.13}
\end{equation}
thus it is required $F''(0)>0$ for stability.

\section{Conclusion}

In this paper we discussed stability conditions for modified teleparallel gravity, which arise from Minkowski stability. It is quite clear that our stability conditions (\ref{2.12}) and (\ref{2.14}) are only necessary but not enough conditions. First of all we discussed only simplest isotropic homogeneous perturbations and taking into account more general types of perturbations may generate additional restrictions. Also there may be other restrictions, which arise from different types of instabilities. Nevertheless, even this simplest analysis gives us essential restrictions for parameters of the theories as we can see from the previous section.

\section{Acknowledgments}

This work was  supported by the RSF grant 16-12-10401.

\end{document}